\documentclass[twocolumn,showpacs,amsmath,amssymb,prl]{revtex4} 
\usepackage{graphicx}

\begin{document}
\title{Reduction of Guided Acoustic Wave Brillouin Scattering in Photonic Crystal Fibers}
\author{D.~Elser}
\email{delser@optik.uni-erlangen.de}
\author{U.~L.~Andersen}
\author{A.~Korn}
\author{O.~Gl\"ockl}
\author{S.~Lorenz}
\author{Ch.~Marquardt}
\author{G.~Leuchs}
\affiliation{Institute of Optics, Information and Photonics (Max Planck Research Group), University of Erlangen-Nuremberg, G\"unther-Scharowsky-Str.~1, Building~24, 91058~Erlangen, Germany}

\date{\today}

\begin{abstract}
Guided Acoustic Wave Brillouin Scattering (GAWBS) generates phase and polarization noise of light propagating in glass fibers. This excess noise affects the performance of various experiments operating at the quantum noise limit. We experimentally demonstrate the reduction of GAWBS noise in a photonic crystal fiber in a broad frequency range using cavity sound dynamics. 
We compare the noise spectrum to the one of a standard fiber and observe a 10-fold noise reduction in the frequency range up to 200~MHz. Based on our measurement results as well as on numerical simulations we establish a model for the reduction of GAWBS noise in photonic crystal fibers.
\end{abstract}


\maketitle

The impact of thermally excited acoustic phonon modes is a severe hindrance for the performance of a variety of physical systems. In gravitational wave detection, for example, the thermal excitation of internal acoustic modes of the interferometer mirrors is a limiting factor to the measurement sensitivity~\cite{Saulson90}, in solid state quantum dots electron-phonon interactions lead to decoherence~\cite{Qin01}, and also in fiber optics the detrimental character of acoustic phonon modes has been encountered~\cite{Shelby85b}.
A general strategy to reduce the amplitude of phonon modes is to cool the environment to extremely low temperatures. This technique is however very inconvenient, in particular when employed to fibers connecting distant points.
Another possibility is making use of cavity sound dynamics and modifying the phonon spectrum by tailoring the boundary conditions for sound waves through spatial structuring.
This has been discussed e.g. in connection with quantum dots: in order to reduce the electron-phonon interaction the acoustic mode spectra have been modified by suspended phonon cavities~\cite{Hoehberger03} and control of electron dephasing in an electron-phonon cavity has been proposed~\cite{Debald02}.
Phonon cavities allow for coherent coupling of discrete electronic states with discrete phonon states~\cite{Weig04}, in resemblance to cavity QED~\cite{Imamoglu99}.
The possibility to completely suppress phonon scattering of excitons by combining different effects, one of them being the fabrication of smaller structures, has been pointed out in~\cite{Nojima92}.
Sound attenuation and phononic band-gaps  have been studied theoretically and experimentally in one-~\cite{Diez00}, two-~\cite{Montero98} and three-~\cite{Liu00} dimensional periodic structures.
Sonic band-gaps in a preform of a photonic crystal fiber (PCF)~\cite{Russell03} have been calculated and measured in~\cite{Russell03b}.
Light-sound interactions can on the other hand be deliberately enhanced by structuring the environment of the phonons, e.g. in a double-cavity being simultaneously resonant for photons and phonons~\cite{Trigo03}.
Recently, confinement of acoustic modes in the core of a photonic crystal fiber has been demonstrated~\cite{Dainese06}.

\begin{figure}
	\centering
		\includegraphics[width=0.5\columnwidth]{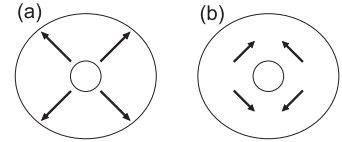}
	\caption{(a) The radial modes $\textrm{R}_{0,m}$ generate phase noise by strain induced modulation of the refractive index. (b) For the same reason, the mixed torsional-radial modes $\textrm{TR}_{2,m}$ also induce phase noise. They additionally produce polarization noise by causing birefringence fluctuations in the fiber.}
	\label{fig:Modes}
\end{figure}

In this Letter we suggest and experimentally demonstrate a convenient method to reduce acoustic modes in optical fibers: by using an appropriately tailored photonic crystal fiber we find that the impact of acoustic vibrations in comparison to a standard fiber can be reduced owing to the microstructure of the light-guiding core.
Thermally excited mechanical vibrations in glass fibers give rise to the so-called Guided Acoustic Wave Brillouin Scattering (GAWBS). Photons propagating in the fiber interact with transverse acoustic phonons and thus experience modifications in phase and polarization. Due to the guiding properties of a glass fiber, photons can be scattered in the forward direction which is not possible in bulk media. 
This mechanism was first discovered by Shelby et al.~\cite{Shelby85b} and has since then been widely studied, especially with regard to its negative impact on the generation of squeezed states~\cite{Shelby86,Poustie92,Bergman92,Poustie93}.
The thermally excited discrete phonon modes are determined by the fiber geometry~\cite{Shelby85b}.
For two types of these modes the overlap between the light field and the acoustic modes is non-vanishing: the radial modes $\textrm{R}_{0,m}$ and the mixed radial-torsional modes $\textrm{TR}_{2,m}$ produce phase modulations of the light via strain induced refractive index changes.  The $\textrm{TR}_{2,m}$-modes additionally depolarize the light via induced birefringence fluctuations (see Fig.~\ref{fig:Modes}). These modes are also referred to as polarized and depolarized GAWBS, respectively.

GAWBS noise reduction is an important issue for squeezing experiments making use of the Kerr nonlinearity of fibers~\cite{Schmitt98,Heersink05}.
The generation of excess noise has an adverse impact on many quantum information experiments relying on squeezed states.
For applications such as entanglement swapping~\cite{vanLoock99,Gloeckl03}, dense coding~\cite{Braunstein00} or quantum cryptography~\cite{Ralph99} a quantum state with a high degree of purity is desired. The excess noise due to GAWBS in a fiber degrades the performance of these applications and in some cases renders them impossible.
Therefore, various efforts have been undertaken to reduce or to compensate for GAWBS noise. Besides cooling the fiber~\cite{Perlmutter90} the influence of GAWBS has been diminished by removing the fiber jacket (thus increasing the quality factor of the phonon modes) or by subtracting the correlated phase noise accumulated by two consecutive pulses having propagated through the same fiber~\cite{Bergman93,Townsend95}. All of these approaches are experimentally challenging. 
In this Letter we show that GAWBS noise can be suppressed by a more convenient method, namely by using a photonic crystal fiber which also acts as a phononic crystal~\cite{Russell03b}.
In a PCF, the core is to some extent mechanically isolated from the cladding since in the structured area the speed of sound is modified and phonon resonances are induced. We therefore expect the coupling between the vibrational cladding modes and the light field in the core to be reduced in comparison to a standard fiber, where core and cladding are strongly coupled. 
Another consequence of a hole structure is the modification of the acoustic velocities which has been investigated recently \cite{Shibata06}.

\begin{figure}
	\centering
\includegraphics[width=0.80\columnwidth]{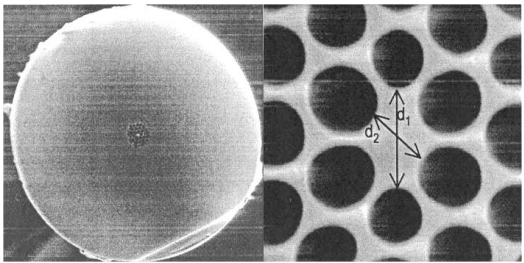}
		\caption{Cross section and close-up of the photonic crystal fiber NL-PM-700. The fiber is polarization maintaining due to an elliptical core ($d_1 \approx 2.4\:\mu \textrm{m}$, $d_2 \approx 1.5\:\mu \textrm{m}$). The pitch between the holes is approx. $1.6\:\mu \textrm{m}$, pitch to hole size ratio~$\approx 0.8$, diameter of the holey region~$\approx 10\:\mu \textrm{m}$, cladding diameter~$\approx~127\:\mu \textrm{m}$ ~\cite{DataSheetNLPM700}.}
	\label{fig:NL-PM-700}
\end{figure}

The internal structure of the PCF under investigation is shown in Fig.~\ref{fig:NL-PM-700}.
We compare the GAWBS characteristics of this PCF with the polarization maintaining standard fiber HB800G from Fibercore, which has a cladding diameter of $80\: \mu \textrm{m}$ and a mode field diameter of $4.5 \: \mu \textrm{m}$.
By always using the same experimental parameters (fiber length, light power, interferometric visibility) we make sure that the measurements are directly comparable.

\begin{figure}
	\centering	\includegraphics[width=0.90\columnwidth]{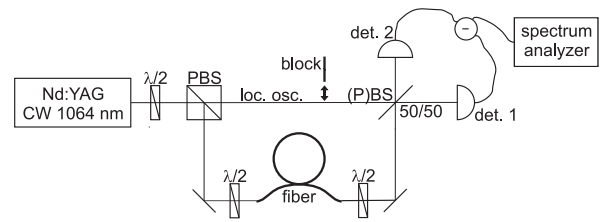}
	\caption{The setup for phase and polarization noise noise measurements. For phase measurements, the local oscillator was unblocked and the optical phase between the two arms of the Mach Zehnder interferometer was locked to $\pi$/2. This results in two equally bright beams at the output of the 50/50 non-polarizing beam splitter.
The visibility in our experiments was 70\%.
For polarization measurements, we blocked the local oscillator and replaced the 50/50 beam splitter by a polarizing beam splitter.
The polarization extinction ratio was~ 2\%.
All measured fibers had a length of $8\:\textrm{m}$ and the light power incident on each detector was approx. $0.9\:\textrm{mW}$.}
	\label{fig:Setup}
\end{figure}

Figure~\ref{fig:Setup} shows our experimental setup for carrying out phase and polarization noise measurements in fibers.
The light source is a continuous-wave single-frequency diode-pumped Nd:YAG nonplanar ring oscillator (Innolight Mephisto 500), which we measured to be shot noise limited for frequencies larger than $10\:\textrm{MHz}$. 
For phase measurements, the local oscillator is unblocked and the setup represents a Mach Zehnder inerferometer.
The half wave plate at the fiber input rotates the linear polarization in order to inject light along the main axis of the polarization maintaining fiber.
The half wave plate after the fiber adjusts the direction of the linear polarization to the local oscillator.
Control over the polarization state is crucial in this experiment in order to obtain a large interferometric visibility. The output from the fiber interferes with the local oscillator of equal intensity on a 50/50 beam splitter and the interference phase is locked to $\pi/2$ phase difference with an active feedback loop.
The difference of the ac signals of the two detectors (New Focus 1611) is fed to a spectrum analyzer.
We determine the shot noise level by blocking the light beam from the fiber and increasing the intensity of the local oscillator in order to obtain the same power on the two detectors as in the interferometric measurement. 
Since the local oscillator is quantum noise limited above $10\:\textrm{MHz}$ its noise properties are known. The contribution of the fiber phase noise can be inferred by subtracting the noise power from the local oscillator.
Under the assumption of a shot noise limited local oscillator, the phase noise is given by the ratio of the variances of the difference signal ($V_{-}$) and the shot noise ($V_\textrm{SN}$):
\begin{equation}
	\frac{V_{-}(\omega)}{V_{\textrm{SN}}(\omega)} = \frac{1}{2} \left( \left\langle \delta \hat{Y}^2(\omega) \right\rangle +1 \right),
\label{eq:phasenoise}
\end{equation}
where
$\omega$ is the measurement frequency (frequency of the acoustic modes) and
$\langle \delta \hat{Y}^2(\omega) \rangle$ is the variance of the phase quadrature in the beam coming from the fiber. 
Equation~(\ref{eq:phasenoise}) holds in the case of a large classical amplitude (bright beams) where only fluctuating terms of first order have to be considered.
We see that if the phase quadrature is at the shot noise limit ($\langle \delta \hat{Y}^2 \rangle = 1$) the ratio equals one. Excess phase noise results in an increased ratio.
For polarization noise measurements, we block the local oscillator and replace the 50/50 beam splitter by a polarizing beam splitter.
The laser beam is again injected onto one axis of the polarization maintaining fiber. This axis now defines the polarization of the bright local oscillator which in this case travels in the fiber. Depolarizing GAWBS scatters photons into the orthogonal polarization. These scattered photons are probed via a homodyne Stokes measurement where the polarizing beam splitter divides the beam into two parts of equal brightness.
Polarization fluctuations become visible in the  difference current of the two detectors.
A high polarization contrast is easily obtained in this setup since no spatial mode-matching is necessary.

\begin{figure}
	\centering
\includegraphics[width=1.00\columnwidth]{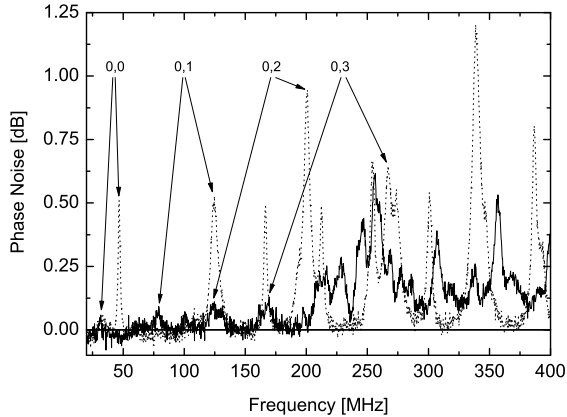}
	\caption{Phase noise accumulated by a light beam of $0.9\:\textrm{mW}$ having passed through $8\:\textrm{m}$ of fiber (solid line: PCF, dotted line: standard fiber). The plots are measured with an interferometric visibility of 70\% and a detection efficiency of 86\%. The small inset numbers (0,$m$) mark the cladding $\textrm{R}_{0,m}$-modes visible in the PCF spectrum as well as the corresponding modes in the standard fiber. The difference of the fiber diameter (standard fiber: $80\:\mu \textrm{m}$, PCF: $127\:\mu \textrm{m}$) leads to frequency shifts between standard fiber and PCF.}
	\label{fig:PhaseNoise}
\end{figure}

\begin{figure}
	\centering
\includegraphics[width=1.00\columnwidth]{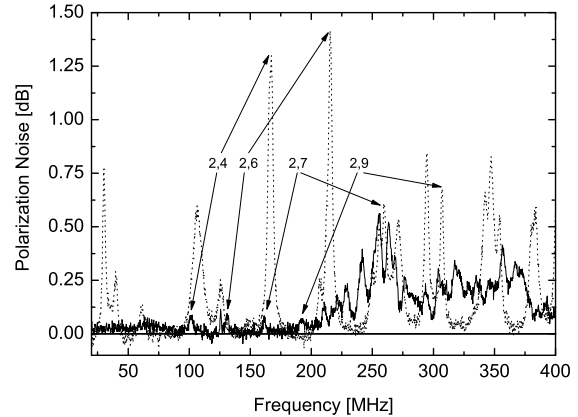}
	\caption{Polarization noise accumulated by a light beam of $1.8\:\textrm{mW}$ having passed through $8\:\textrm{m}$ of fiber (solid line: PCF, dotted line: standard fiber). The plots are measured with an polarization extinction ratio of 2\% and a detection efficiency of 86\%.
The small inset numbers (2,$m$) mark the cladding $\textrm{TR}_{2,m}$-modes visible in the PCF spectrum as well as the corresponding modes in the standard fiber. 
}
	\label{fig:PolarizationNoise}
\end{figure}

The GAWBS noise spectra obtained by our measurements are shown in Fig.~\ref{fig:PhaseNoise} (phase noise) and Fig.~\ref{fig:PolarizationNoise} (polarization noise). The traces were recorded using a spectrum analyzer at a resolution bandwidth of $1\:\textrm{MHz}$, a video bandwidth of $30\:\textrm{Hz}$ and a ten times averaging. Since we use relatively short fibers all measured peaks are rather weak. The coating had not been removed from the fiber, resulting in a broad linewidth, as the acoustic modes are damped in the plastic~\cite{Hardman96}.
We clearly observe a strong noise reduction in the frequency range up to $200\:\textrm{MHz}$: the phase noise is suppressed by a factor of roughly~7 while the polarization noise is reduced by a factor of more than~13.

A two-dimensional model where the vibrational modes of a long cylinder are considered gives the frequencies and amplitudes of the two sets of transverse modes~\cite{Shelby85b}.
By applying this model we calculate the theoretically expected frequencies of the GAWBS modes. In the case of the standard fiber, we thereby can attribute the measured peaks to the cladding modes $\textrm{R}_{0,m}$ ($m$=0,..,4) and $\textrm{TR}_{2,m}$ ($m$=0,..,12). As in \cite{Shelby85b}, the mode $\textrm{TR}_{2,5}$ is not observable in the experimental data. 
Some peaks appear additionally to the theoretically predicted. We assume these to stem from geometrical asymmetries due to the polarization maintaining property of the fiber or the details of the phononic structured area.

\begin{figure}
  \centering
  \begin{minipage}[b]{0.49\columnwidth}
    \includegraphics[width=\columnwidth]{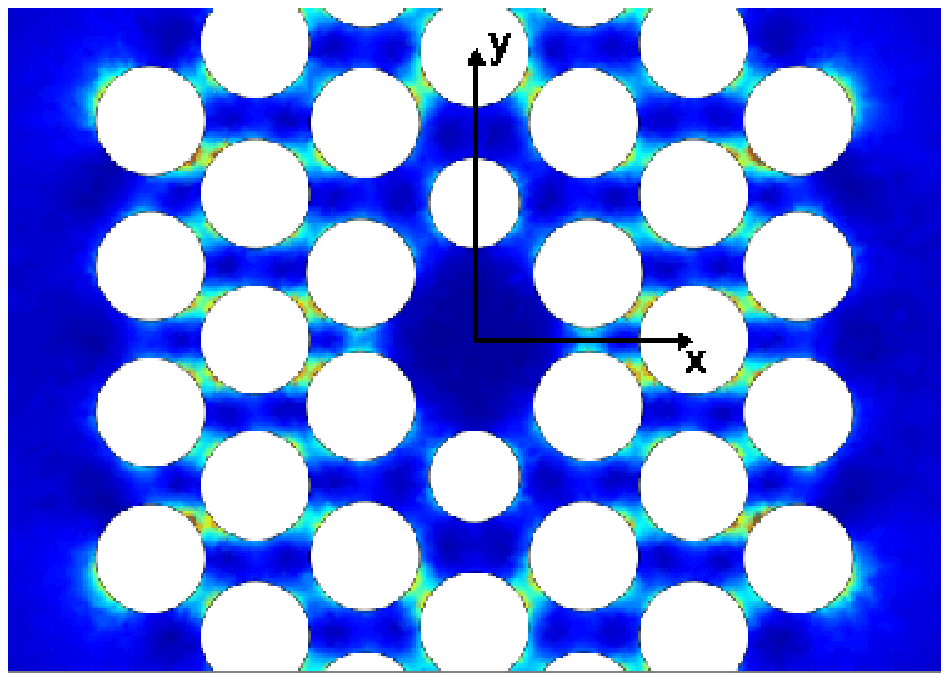}  
  \end{minipage}
  \begin{minipage}[b]{0.49\columnwidth}
    \includegraphics[width=\columnwidth]{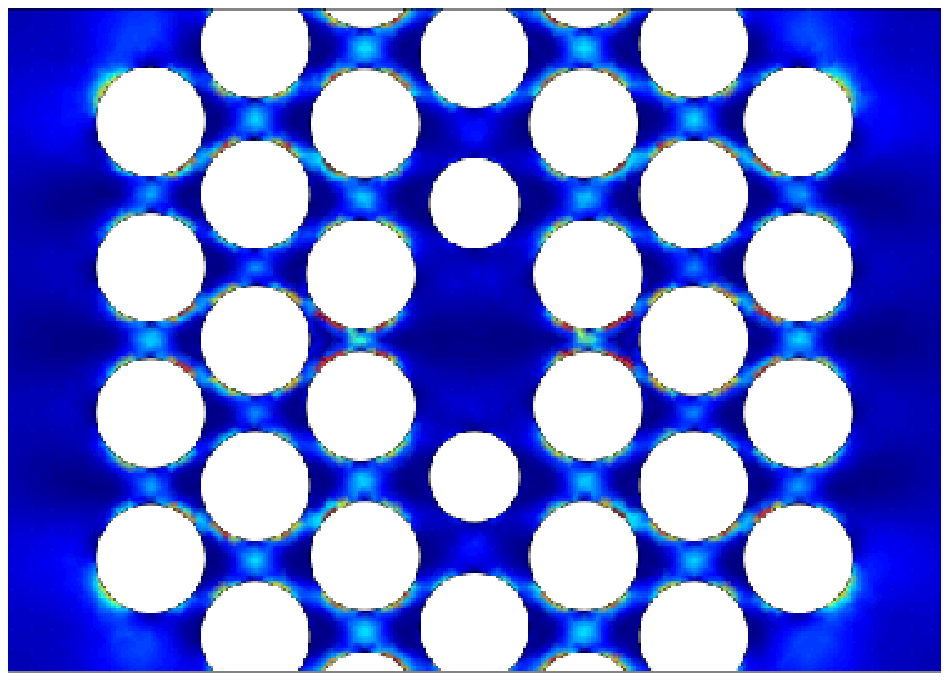}  
  \end{minipage}
  \caption{Strain energy density for the fundamental modes $\textrm{R}_{0,0}$ (left) and $\textrm{TR}_{2,0}$ (right) in the PCF (blue = low energy, red = high energy). The hole structure prevents the strain field from entering into the light-guiding core. Below $200\:\textrm{MHz}$, a similar behavior is found for the higher order $\textrm{R}_{0,m}$- and $\textrm{TR}_{2,m}$-modes.} 
  \label{fig:pcf_modes}
\end{figure}

\begin{figure}
		\centering
\begin{minipage}[b]{0.49\columnwidth}
    \includegraphics[width=\columnwidth]{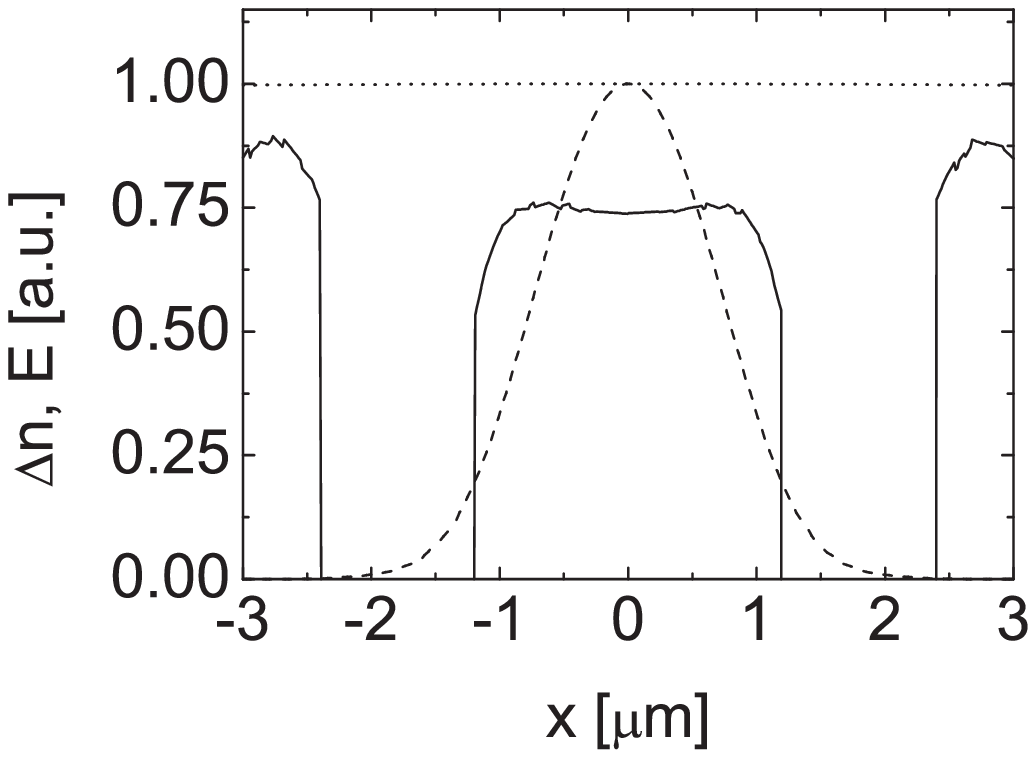}  
  \end{minipage}
  \begin{minipage}[b]{0.49\columnwidth}
    \includegraphics[width=\columnwidth]{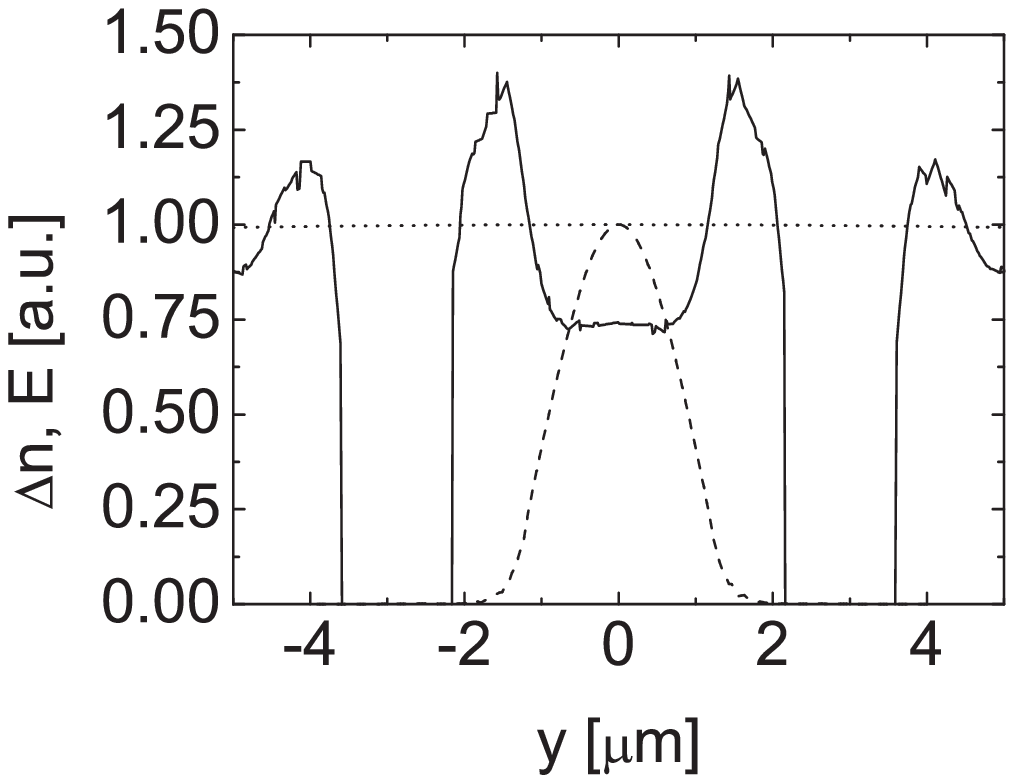}  
  \end{minipage}
	\caption{Cross section plots of the refractive index modulation of the $\textrm{R}_{0,0}$ mode in the PCF (solid line), in comparison to a standard fiber with the same diameter (dotted line). The dashed curve represents the electrical field amplitude of the light field.
	$x$ and $y$ refer to the horizontal and vertical cuts through the fiber cross section (Fig.~\ref{fig:pcf_modes}).}
	\label{fig:refr_index_modulation}
\end{figure}

In case of the PCF, some of the solid cylinder modes, namely $\textrm{R}_{0,m}$ ($m=0,\dots,3$) and $\textrm{TR}_{2,m}$ ($m=4, 6, 7, 9$), appear in the spectrum in the frequency range below $200\:\textrm{MHz}$. It is clearly visible that the amplitudes of these modes are strongly attenuated in comparison to the standard fiber.
Hence we conclude that the cladding modes couple only weakly to the light-guiding core. 
We confirm this conclusion by numerical finite elements simulations using the software package `Comsol Multiphysics' (Femlab): We perform an eigenfrequency analysis in a 2-dimensional plane strain model assuming a free fiber surface (no constrains on boundaries).
The simulations give us the displacement vector $\textbf{u}$ for each vertex of the finite elements mesh. 
From the displacement we calculate the strain energy density, depicted in Fig.~\ref{fig:pcf_modes} for the two fundamental modes ($\textrm{R}_{0,0}$ and $\textrm{TR}_{2,0}$).
We clearly see that due to the hole structure the strain field cannot penetrate into the light-guiding core. Elasto-optic index and birefringence modulations are thus reduced which leads to decreased GAWBS noise.
From the numerical simulations we also find the refractive index modulation $\Delta n$ as shown in Fig.~\ref{fig:refr_index_modulation} for the $\textrm{R}_{0,0}$~\footnote{For radial modes, $\Delta n$ is proportional to the divergence of the displacement $\textbf{u}$.}.
We find a redistribution of the index modulation due to the hole structure, and we clearly see that the light-guiding core is less affected by the modulation compared to the standard fiber (dotted curve).
The phase noise power of light propagating in the fiber can be determined by averaging the index modulation over the mode profile of the electrical field amplitude (dashed curve).

Describing the acoustic modes in the PCF by the standard modes of a solid cylinder is only possible as long as the cylinder modes are not essentially perturbed by the hole structure (apart from the damping of the vibrations in the core). At frequencies around $200\:\textrm{MHz}$, the innermost maximum of the displacement is located near the hole structure.
Thereby, the simulations show a multitude of new modes, not included in the cylinder model. These modes lead to a quasi-continuous noise spectrum above $200\:\textrm{MHz}$, as evident from Fig.~\ref{fig:PhaseNoise} and~\ref{fig:PolarizationNoise}.

The core itself can be excited by acoustic vibrations at frequencies above $1\:\textrm{GHz}$, which are beyond our measured range. Ultra-efficient acoustooptic devices could be built making use of the sound confinement for these core modes~\cite{Dainese06}.



In conclusion, we have found that the investigated photonic crystal fiber offers the advantage of reducing GAWBS noise in a wide frequency range. We have shown that the light propagating in the PCF does not sample a significant amount of the acoustic vibrations supported by the cladding structure. 
For squeezing experiments, the wide range of design parameters in PCFs offers a twofold advantage: besides the possibility to suppress GAWBS noise in a desired frequency range, the attainable high non-linearities potentially   allow for utilizing very short fibers, additionally reducing the excess noise.
The intensity of GAWBS modes could on the other hand be deliberately enhanced by a particular PCF design, e.g. in order to build sensors for temperature~\cite{Tanaka04} or tensile strain~\cite{Tanaka99}, as well as acousto-optic modulators \cite{Dainese06}.

The authors acknowledge support by the Deut\-sche For\-schungs\-ge\-mein\-schaft (FG532) and by the \mbox{COVAQIAL} project (no. FP6-511004).















\end{document}